\documentclass[preprint,11pt,number]{elsarticle}

\usepackage{amssymb}
\usepackage{graphicx}
\usepackage{amsmath}
\usepackage{multirow}
\usepackage{nomencl}
\usepackage{url}
\makenomenclature

\journal{ }
\usepackage{etoolbox}
\renewcommand\nomgroup[1]{%
  \item[\bfseries
  \ifstrequal{#1}{A}{Acronyms}{%
  \ifstrequal{#1}{I}{Indices}{%
  \ifstrequal{#1}{P}{Parameters}{%
  \ifstrequal{#1}{C}{Continuous variables}{%
  \ifstrequal{#1}{B}{Binary variables}{}}}}}%
]}

\begin{document}

\begin{frontmatter}



\title{Renewable energy communities: do they have a business case in Flanders?}


\author[inst1]{Alex Felice}
\author[inst2]{Lucija Rakocevic}
\author[inst2]{Leen Peeters}
\author[inst1]{Maarten Messagie}
\author[inst1]{Thierry Coosemans}
\author[inst1]{Luis Ramirez Camargo}
\affiliation[inst1]{organization={Electric Vehicle and Energy Research Group (EVERGI), Mobility, Logistics and Automotive Technology Research Centre (MOBI), Department of Electrical Engineering and Energy Technology, Vrije Universiteit Brussel},
            city={1050 Ixelles},
            country={Belgium}}
\affiliation[inst2]{organization={Th!nk E},
            addressline={Philipssite 5}, 
            city={3001 Leuven},
            country={Belgium}}

\begin{abstract}
Renewable energy communities (RECs) are prominent initiatives to provide end consumers an active role in the energy sector, raise awareness on the importance of renewable energy (RE) technologies and increase their share in the energy system thus reducing greenhouse gas emissions. The economic viability of RECs though, depends on multiple interdependent factors that require careful examination for each individual context. This study aims at investigating the impact of electricity tariffs, ratio of electrification of heating and transportation sectors, prices of RE technologies and storage systems, and internal electricity exchange prices on the annual cost for electricity provision of a REC. A mixed-integer linear model is developed to minimize energy provision costs for a representative REC in Flanders, Belgium. The results indicate that RECs have the potential to reduce these costs by 10 to 26\% compared to business-as-usual. This cost reduction depends on the type of electricity tariffs and the level of uptake of flexible assets such as heat pumps and electric vehicles. The shift towards a higher power component in the electricity tariff makes electricity storage systems more attractive, which leads to higher electricity self-consumption. The introduction of flexible assets adds the possibility to shift demand when tariffs are lower and makes larger sizes of photovoltaic systems economically viable due to the increase in the total electricity demand. However, RECs cost reduction compared to individual smart-homes amounts to only 4\% - 6\% in the best cases. Uncertainties stemming from the regulation and the costs of setting up a REC may reduce the estimated benefits.
\end{abstract}




\end{frontmatter}

\section{Introduction}
\label{sec:introduction}
Renewable energy communities (RECs) are indicated as one of the means to help democratize, decarbonize and decentralize the energy sector across Europe. As defined in the recast of the European Renewable Energy Directive (RED II) \cite{DirectiveEU2018/2001oftheEuropeanParliamentandoftheCouncil2018}, a REC is a legal entity entitled to produce, consume, store and sell renewable energy between geographically co-located private citizens, public entities and SMEs. Their objectives are to create economical, environmental and social benefits to the community members, as well as to increase local acceptance of renewable energy projects \cite{Alaton2020}. Although the RED II provides a framework for the implementation of RECs, the specific conditions for each country depend on the transposition at the individual EU member state level, meaning that there are still uncertainties on the conditions for an extensive uptake of RECs \cite{Hoicka2021}. Ines et al. \cite{Ines2020} highlight the transposition problem by comparing regulations of nine different European countries or regions. They identified that the first challenge for RECs implementation is to overcome local legal barriers in order to exploit the opportunities brought by the legal framework at EU level. According to Brummer et al.\citep{Brummer2018}, this dependency on regulations cuts both ways: regulation promoting RECs may be fruitful for their uptake, but it might present a weakness for long term development of RECs.\par

When the regulation allows the development of RECs, the next question is to understand the motivation of citizens to join a REC. For instance, Conradie et al.\cite{Conradie2021} focus on better understanding the factors that influence members' participation in a community in Flanders, Belgium. They showed that lowering the practical barriers of entry in a REC are not sufficient alone. Attitude towards renewable energy sources (RES), ecological impact and expected financial gains are also motivators. According to Bauwens et al.\cite{Bauwens2018}, acceptance of new RES projects is higher for RECs' members than for non-members, highlighting their social impact. The importance of the economical benefits, which will determine how much investment will be done in new community renewable energy projects, is analyzed in another work of Bauwens et al. \cite{Bauwens2019}: results show that the return on investment is the most important determinant for members of large communities of interest, while environmental, social and other non-economic drivers tend to dominate financial motives for members of smaller communities of place. This result is also in accord with the RED II \cite{DirectiveEU2018/2001oftheEuropeanParliamentandoftheCouncil2018}: the main objective of a REC should not be a pure financial gain.
In this study, we propose a MILP optimization model where we incorporate different non-technical factors, such as tariffs or investment's strategies, that could influence RECs performance. This is done by creating 156 different scenarios and analyzing their impact on the final electricity costs for the users, as well on uptake of PV and BESS. Aside from final annual costs, we use self-consumption and self-sufficiency measures as indicators of performance.\par

\section{Literature review}
A REC can be viewed as economically viable when the total energy cost for the community members are at parity with or lower than other options for energy supply. Community members are assumed to be consumers or prosumers. Electricity tariff directly affects the economic viability of a REC. Therefore a detailed analysis of both technical and non-technical aspects of REC are needed to understand its economic viability. Radl et al. \cite{Radl2020} compared photovoltaic (PV) and battery energy storage systems (BESS) profitability in multi-energy RECs for eight different European countries. The authors concluded that except for cases of full-load hours dictated by weather conditions, the electricity tariffs has the highest impact on PV investments. Concerning BESS, they concluded that under current market conditions they are rarely profitable except when capacity based pricing is applied. \par

Integral part of electricity tariff is the network tariff. New network tariff structures may also impact the REC business case. Traditionally, the main part of consumers’ network tariffs are based on their volumetric electricity extraction from the grid, in \texteuro/kWh. With an increasing share of prosumers, who may both extract and inject electricity to the grid, and a growing challenge of managing power peaks in the grid due to more intermittent generation, the traditional volumetric network tariffs have become outdated. The affordability of decentralized RES has led to an increasing number of consumers with an alternative energy supply and thus an ability to react and momentarily opt out of the energy supply from the grid. Tariffs for prosumers and energy communities need to reflect how these types of consumers now have an alternative energy supply but remain connected and still dependent on the grid, and that power flows go two ways \cite{Fridgen2018}. There is a consensus that volumetric tariffs with net-metering are unfit for the future high-RES energy system (e.g. \cite{Ansarin2020}; \cite{Clastres2019}; \cite{Schittekatte2018}). The process of reviewing tariff structures has therefore been initiated in many countries. Abada et al. \cite{Abada2020} studied the impact of electricity tariff design on energy community formation. They find that fixed tariffs lead to REC formation while also generating the most social welfare and avoiding over-investments. Capacity based and volumetric tariffs incentivize RES investments, but may also lead to a welfare destructive snowball effect of over-investments. The impact of potential future tariff structures on the business case for REC is however not well understood. It is important to better understand the interaction between the new tariffs and the promotion of REC, in order to streamline the policy initiatives.\par 
Another important point mentioned on the RED II \cite{DirectiveEU2018/2001oftheEuropeanParliamentandoftheCouncil2018} is the use of renewable electricity production in heating and cooling and transport sectors for reducing greenhouse gas emissions and fuel dependency. At residential level this can be effectively achieved by increasing the usage of technologies like heat pumps (HPs) and electric vehicles (EVs). The electrification of heating and cooling and transportation sectors at the building and neighbourhood level introduces the need for appropriate techno-economic models for multi-energy systems in order to identify optimal investments and operational strategies. However, the challenge for modelling RECs is not only technical but also policy dependent. This  increases the complexity and the computational resources needed for such techno-economic models to be used. The literature covering optimization of RECs, or district-level multi-energy systems in general, is already extensive, but the inclusion of the impact of different policy and/or investment possibilities is still under-represented. Weckesser at al. \cite{Weckesser2021} introduced the regulatory aspects by analyzing different community configurations on the distribution grid, while optimizing the size of PV and battery storage of a REC for minimizing costs and disturbance on the low-voltage distribution grid. In \cite{Fioriti2021} the optimal sizing and operation of energy communities is coupled with a study of a business model for the participation of aggregators in a REC while ensuring fair sharing of costs and revenues between all the actors. Braeuer et al. \cite{Braeuer2022} applied the German Tenant Electricity Law, a particular regulation in place between tenants and owners of multi-apartment buildings, to a mixed-integer linear program (MILP) optimisation model for an energy community composed of multi-apartment buildings. The results show how the legal framework has a direct impact the economic viability of the REC. Another analysis regarding multi-apartment buildings is presented in \cite{Fina2018}, where the difference in legislative framework between Austria and Germany results in different profitability of shared PV systems in multi-apartment buildings. Due to policy differences, the profitability of such systems in Austria is very marginal compared to the Germany. A policy-oriented optimization framework was developed in \cite{Scheller2018}, further highlighting the needs of merging techno-economic aspects with regulatory ones. A previous study of the authors \cite{Felice2021} investigates the conditions needed by RECs to operate in an economic positive way in the context of Flanders. Results indicate that even though user type, user consumption and electricity tariffs are important factors, the amount of flexible technology in a REC is the most important factor to reduce operational costs.\par
The examples given above show that there is ongoing work on the coupling of regulatory and techno-economic aspect in energy modelling, but these studies are bound to regional and/or national levels. The novelty of this work relies on the extensive scenarios analysis to simultaneously map the impact of most uncertainties on the final cost for energy provision for the users and renewable penetration by the means of RECs. To the best knowledge of the authors this is the first study tackling these subjects for the case of Flanders, and due to the expected role of RECs for the energy transition plan in the EU it can contribute to inspire similar analysis for other regions or countries.\par
\section{Methods}
\label{sec:methods}
\subsection{REC set-up}
\label{sec:study}
The REC set up is the result of a participatory process between multiple stakeholders. It is a synthetic REC composed by eleven real residential buildings, located in Flanders, with their associated hourly electrical consumption profiles for a whole year. These profiles are all provided by Fluvius, the Flemish distribution system operator. Between these eleven members, nine of them are single-family houses while the last two are apartment buildings. All of them are connected to the low-voltage grid only. The Distributed energy resources (DER) included in the system are PV, BESS, EV chargers and controllable HP. PV output profiles are calculated by using a single normalized generation profile for Flanders: 1,000 kWh of energy produced in a year for 1 kWp installed, which is scaled with the different capacities installed for every member. EV chargers demand profiles are based on a fixed daily demand of 7 kWh. Two type of residential heating profiles are simulated in TRNSYS \cite{TRNSYS2019} based on building type, usage and outside temperature. With the same principle, heat pumps' COP are simulated based on heat demand and outside temperature. The heat demand profiles are then translated in electricity demand of heat pumps using the COP profiles. All the technical and economical parameters are listed in Table \ref{tab:tech-eco-param}. \par
\begin{table}[ht]
\caption{Technical and economical parameters}
\label{tab:tech-eco-param}
\centering
\begin{tabular}{cc}
\hline
\textbf{Parameter} & \textbf{Value} \\
\hline
$\eta^{ch}$/$\eta^{disch}$ & 95 \% \\
$SOC_{min}$ & 0.1 \\
$\overline{p^{ch}}$ & 1 \\
$\overline{p^{disch}}$ & 2 \\
$m^{ev}$ & 3.5 kW \\
$m^{hp1}$ & 1 kW \\
$m^{hp2}$ & 2 kW \\
$lifetime^{PV}$ & 25 years \\
$lifetime^{BESS}$ & 10 years \\
$\lambda^{ext,im}_{t,n}$ & 28 c\texteuro/kWh or 21 c\texteuro/kWh \\
$\lambda^{ext,ex}_{t,n}$ & 3.5 c\texteuro/kWh \\
$\lambda^{p}_{n}$ & 4.17 \texteuro/kW/month \\
$d$ & 7.5 \% \\
\hline
\end{tabular}
\end{table}
\subsection{Scenarios construction}
\label{sec:scenarios}
We create an extensive set of scenarios to evaluate different options of REC set-up for having a positive business-case. The first set of scenarios concerns the presence of HPs, for both heating and cooling, and EV chargers. The penetration of these technologies is varied in order to quantify the impact of different levels of flexible demand present in the system. The second set of scenarios, Capex scenarios, show sensitivity of cost of energy provisions on the investment cost of PV and BESS. The third set of scenarios concerns the electricity tariffs, where a comparison between volumetric and capacity tariff is made.  The fourth set of scenarios aims at comparing three different investment strategies: business-as-usual (no additional RES are installed), individual investment and community investment. In the community investment case an additional level of sensitivity analysis is performed on the internal energy exchange prices.
\subsubsection{Technologies scenarios}
While PV and BESS are variables of the optimal planning problem for all the scenarios, HPs and EV chargers capacities are incorporated as fixed parameters for selected scenarios. This allows us to compare the economic viability of RECs when technologies that provide the services of heating and transportation using electricity, and therefore provide demand-side management (DSM) capabilities, are available or not. Installed capacities of heat pumps are fixed based on the maximum daily demand that needs to be satisfied, while EV chargers' maximum capacities are standards present in the LV grid. In total four technology scenarios where created to assess the impact of different degree of flexible demand present in the system. These are summarized in Table \ref{tab:tech-scenarios}.
\begin{center}
\begin{table}[ht]
\caption{Technology scenarios}
\label{tab:tech-scenarios}
\centering
\begin{tabular}{cccc}
\hline
\textbf{Scenario} & \textbf{\begin{tabular}[c]{@{}c@{}}\% of members\\ with HP\end{tabular}} & \textbf{\begin{tabular}[c]{@{}c@{}}\% of member with\\ EV charger\end{tabular}} & \textbf{\begin{tabular}[c]{@{}c@{}}\% of flexible\\ demand\end{tabular}} \\
\hline
\textbf{1} & 100 & 100 & 41.8 \\
\textbf{2} & 100 & 0 & 25.6 \\
\textbf{3} & 0 & 100 & 26.9 \\
\textbf{4} & 0 & 0 & 0 \\
\hline
\end{tabular}
\end{table}
\end{center}
\subsubsection{Capex scenarios}
As a techno-economic optimization model is used, economic parameters like the investment cost for newly installed PV and BESS play a fundamental role on the outcomes of the model. In order to analyze their impact three different prices are proposed for each technology, starting from a higher price which reflect actual cost of installations and ending with a lower price which is expected in the near future, or results from use of government's subsidies. These parameters were defined in iterative consultation with local technology providers, DSOs and research institutions working on RECs in the Flemish context. The three scenarios are summarized in Table \ref{tab:capex-scenarios}.
\begin{center}
\begin{table}[ht]
\caption{Capex scenario}
\label{tab:capex-scenarios}
\centering
\begin{tabular}{ccc}
\hline
\textbf{Scenario} & \textbf{\begin{tabular}[c]{@{}c@{}}PV installation\\ cost\end{tabular}} & \textbf{\begin{tabular}[c]{@{}c@{}}BESS installations\\ cost \end{tabular}} \\
\hline
\textbf{1} & 1200 \texteuro/kWp & 1000 \texteuro/kWh  \\
\textbf{2} & 1000 \texteuro/kWp & 750 \texteuro/kWh  \\
\textbf{3} & 800 \texteuro/kWp & 500 \texteuro/kWh \\
\hline
\end{tabular}
\end{table}
\end{center}
\subsubsection{Tariff scenarios}
Three scenarios are proposed to analyze the difference in the total annual costs and newly installed DER. We compare a common volumetric tariff to two different capacity tariffs. The reference tariff used for this study is a volumetric one (\texteuro/kWh) with peak and off-peak tariffs. This choice has been made in order to exploit the DSM potential of the system, which cannot be done if a flat tariff is used. Peak times are between 7:00 and 22:00 during weekdays, while off-peak times are the rest of the day in weekdays and during weekends. The two capacity-based tariffs are built by first identifying the final cost split of an electricity tariff in Flanders. From \cite{VREG2020}, a report of the Flemish Regulator of the Electricity and Gas Market (VREG), we have the following structure: 28 \% is the commodity part, 18 \% the DSO tariff, 7 \% the TSO tariff. 17 \% the VAT and 30 \% are fees and taxes. The first capacity tariff represent the planned scenario for 2022 in Flanders \cite{VREG2020} where the DSO tariff will be billed based on the highest monthly peak consumption (\texteuro/kW). For the second capacity tariff all the components except VAT, fees and taxes are kW-based, hence 53 \% of the total. In all the scenarios, injection price for over-production is fixed over the whole time-horizon. It has to be noted that the values in Table \ref{tab:tech-eco-param} are used as a reference to build all the different tariff and peer-to-peer scenarios explored in this study. The import prices $\lambda^{ext,im}_{t,n}$ (average value for Flanders in 2020 \cite{Eurostat2020}) represent the purely volumetric tariff presented in Table \ref{tab:tar-scenarios}, these values are then scaled down to be used for the other tariff scenarios. The same apply for the peak import price $\lambda^{p}_{n}$, where the value in Table \ref{tab:tech-eco-param} is the reference value for tariff scenario 2 which is scaled up using the percentage of tariff scenario 3.
\begin{center}
\begin{table}[ht]
\caption{Tariff scenarios}
\label{tab:tar-scenarios}
\centering
\begin{tabular}{cc}
\hline
\textbf{Scenario} & \textbf{\begin{tabular}[c]{@{}c@{}}Description\end{tabular}} \\
\hline
\textbf{1} & Volumetric \\
\textbf{2} & 18 \% capacity-based \\
\textbf{3} & 53 \% capacity-based \\
\hline
\end{tabular}
\end{table}
\end{center}
\subsubsection{Investment scenarios}
Three investment scenarios are introduced by the mean of three different optimization levels. The first one is the business-as-usual case, where no REC is created and investment in PV and BESS are not introduced. This will be used as a reference case. The second scenario is called individual investment scenario: every member makes investment decision based on own needs, without consideration to other members (individual objective function). In this scenario, no REC is created, energy exchange is allowed only with the grid. The third scenario is the community-joint investment, where the optimal investment in PV and BESS is determined at the REC level and each member can own a share of the assets. A REC is created, meaning that energy can be traded inside the community and with the grid. In this last scenario costs of PV and BESS are assumed to be 10 \% lower than the individual case due to economy of scale. The peer-to-peer exchange is possible in cases where users are not part of the REC, but we include it only in scenario three to assure extreme cases are analyze considering the need to limit the number of scenarios. Such scenario design should already enable to quantify the general impact of different strategies. Another difference between the three optimization levels is in the usage of flexible assets: in the reference scenario, both HPs and EV chargers demand are fixed hourly profiles in order to disable any optimization. While in the individual investment scenario (scenario 2) and in the REC scenario (scenario 3) these two types of profiles are transformed in daily demand to enable hourly optimization of their usage.
\begin{center}
\begin{table}[ht]
\caption{Investment scenarios}
\label{tab:inv-scenarios}
\centering
\begin{tabular}{cc}
\hline
\textbf{Scenario} & \textbf{\begin{tabular}[c]{@{}c@{}}Description\end{tabular}} \\
\hline
\textbf{1} & \begin{tabular}[c]{@{}c@{}}No investment and internal exchange disabled\end{tabular}  \\
\textbf{2} & \begin{tabular}[c]{@{}c@{}}Individual investment and internal exchange disabled\end{tabular}  \\
\textbf{3} & \begin{tabular}[c]{@{}c@{}}Collective investment and internal exchange enabled\end{tabular} \\
\hline
\end{tabular}
\end{table}
\end{center}
\subsubsection{Peer-to-peer scenarios}
These scenarios only apply for the third investment scenario (creation of a REC). For this particular case additional analysis are done on the impact of the peer-to-peer energy price inside the REC. This exchange price is modeled as the difference between the buying and selling price of energy coming from the assets present in the community. Comparison will be made between free internal exchange, an internal cost that correspond to 65 \% (DSO cost + VAT + taxes) and 72 \% (DSO + TSO cost + VAT + taxes) of the buying price from the grid of each tariff scenario, summarized in Table \ref{tab:p2p-scenarios}.
\begin{center}
\begin{table}[ht]
\caption{Internal price scenarios}
\label{tab:p2p-scenarios}
\centering
\begin{tabular}{cc}
\hline
\textbf{Scenario} & \textbf{\begin{tabular}[c]{@{}c@{}}Description\end{tabular}} \\
\hline
\textbf{1} & 0 \% of buying price \\
\textbf{2} & 65 \% of buying price \\
\textbf{3} & 72 \% of buying price \\
\hline
\end{tabular}
\end{table}
\end{center}

\subsection{Key performance indicators}
\label{sec:kpi}
In order to analyze the results, five different KPIs are used:
\begin{itemize}
    \item annualized total cost per kWh consumed (\texteuro/kWh):
    \begin{equation}
    \label{eq:lcoe}    
        \dfrac{C_{inv}+C_{op}}{\sum_{\substack{t\in T,\\n\in N}}(l_{t,n}\ +\ p^{hp}_{t,n}\ +\ p^{cool}_{t,n}\ +\ p^{ev}_{t,n})\cdot \Delta t}
    \end{equation}
    where $C_{inv}$ represents the equivalent annual cost of investment, while $C_{op}$ is the total cost for operating the energy community, both expressed in \texteuro. $l_{t,n}$ is the power demand, $ p^{hp}_{t,n} $ the heat demand, $ p^{cool}_{t,n} $ the cooling demand, $ p^{ev}_{t,n} $ the EV charger demand, all expressed in kW. $\Delta t$ the timestep size in hours.
    \item PV capacity installed per MWh consumed (kWp/MWh):
    \begin{equation}
    \label{eq:pv-kpi}    
        \dfrac{\sum_{n\in N} cap_{PV,n} \cdot 1000}{\sum_{\substack{t\in T,\\n\in N}}(l_{t,n}\ +\ p^{hp}_{t,n}\ +\ p^{cool}_{t,n}\ +\ p^{ev}_{t,n})\cdot \Delta t}
    \end{equation}
    with $cap_{PV,n}$ being the installed PV capacity in kW
    \item BESS capacity installed per MWh consumed (kWp/MWh):
    \begin{equation}
    \label{eq:batt-kpi}    
        \dfrac{\sum_{n\in N} cap_{batt,n} \cdot 1000}{\sum_{\substack{t\in T,\\n\in N}}(l_{t,n}\ +\ p^{hp}_{t,n}\ +\ p^{cool}_{t,n}\ +\ p^{ev}_{t,n})\cdot \Delta t}
    \end{equation}
    with $cap_{batt,n}$ being the installed BESS capacity in kWh
    \item self-consumption ratio (\%):
    \begin{equation}
    \label{eq:scr}
        100 \cdot \dfrac{\sum_{\substack{t\in T,\\n\in N}}p^{pv}_{t,n} - P^{ex}_{t,n}}{\sum_{\substack{t\in T,\\n\in N}}p^{pv}_{t,n}}
    \end{equation}
    where $p^{pv}_{t,n}$ is the power produced by the PV and $P^{ex}_{t,n}$ the power exported to the grid
    \item self-sufficiency ratio (\%):
    \begin{equation}
    \label{eq:ssr}
        100 \cdot \dfrac{\sum_{\substack{t\in T,\\n\in N}}p^{pv}_{t,n} - P^{ex}_{t,n}}{\sum_{\substack{t\in T,\\n\in N}}(l_{t,n}\ +\ p^{hp}_{t,n}\ +\ p^{cool}_{t,n}\ +\ p^{ev}_{t,n})}
    \end{equation}
\end{itemize}
The first three KPIs, including annualized total cost per kWh consumed, PV capacity installed per MWh consumed and BESS capacity installed per MWh consumed, allow the comparison between scenarios that do not have the same total electrical consumption. Self-consumption ratio represents the ratio of energy produced from PV that is used inside the system and not sold. It can be seen as an indicator of over-sizing of PV. Self-sufficiency ratio is instead the ratio between the energy locally produced and consumed and the total energy demand, it is a measure of independence from the main grid.
\subsection{Optimization-problem formulation}
\label{sec:math}
In this section, the optimization problem for finding the minimum annualized cost for energy provision for each REC configuration is presented. The mathematical formulation presented here refers to the third investment scenario (see Section \ref{sec:scenarios}), the community case. In the cases of the first and second investment scenarios part of the equations become simply zero due to the lack of investments or no peer-to-peer exchange. The objective function for the whole time-horizon is
\begin{equation}
\label{eq:obj}
    min\ C_{inv} + C_{op}
\end{equation}
where $C_{inv}$ is the equivalent annual cost of investment, while $C_{op}$ is the total cost for operating the energy community. The equivalent annual cost of investment can be calculated as
\begin{equation}
\label{eq:inv}
    C_{inv} = \sum_{\substack{i\in I,\\n\in N}} y_{i, n} \cdot cap_{i, n} \cdot C_i \cdot CRF_{i}
\end{equation}
with $I$ being the set of all technologies included, $N$ is the set of all REC members. $y_{i,n}$ is a binary variable indicating if the installed capacity of technology $i$ of member $n$, $cap_{i,n}$, is newly installed or was already part of the system. Furthermore, $C_i$ is the cost for the installation of technology $i$ and $CRF_i$ is the calculated capital recovery factor for technology $i$, which is defined by
\begin{equation}
\label{eq:crf}
   CRF_i = \dfrac{d(1+d)^{lifetime_i}}{(1+d)^{lifetime_i}-1}
\end{equation}
where $d$ represents the discount rate and $lifetime_i$ is the lifetime of technology $i$. The operational cost $C_{op}$ introduced in Eq. \ref{eq:obj} is calculated as
\begin{equation}
\begin{split}
    \label{eq:op}
    C_{op} = \sum_{\substack{n\in N,\\t\in T}}\ (P^{im}_{t,n} \cdot \lambda_{t,n}^{ext,im}\ -\ P^{ex}_{t,n}\cdot \lambda_{t,n}^{ext,ex})\cdot \Delta t\ \\ +\  
    \sum_{\substack{n\in N,\\m\in M}}\ (P^{p}_{m,n} \cdot \lambda^{p}_{n})\ +\  \sum_{n \in N}\ C^{int}_{n}
\end{split}
\end{equation}
which is composed by three summation terms. The first one represents the difference between the cost of importing energy and the gain for injecting energy back to the grid. $P^{im}_{t,n}$ and $P^{ex}_{t,n}$ are the power imported from and exported to the main grid for every timestep $t$ and member $n$, while $\lambda^{ext,im}_{t,n}$ and $\lambda^{ext,ex}_{t,n}$ are the tariffs for importing an exporting energy respectively. In order to link power values with cost per energy, we introduce $\Delta t$ as the difference between two timesteps in hours. The second term is the sum of peak consumption cost for every month of the year, with $P^{p}_{m,n}$ being the peak power imported from the grid for month $m$ and member $n$ and $\lambda^{p}_{n}$ is its associated cost. Finally, the last summation term is the the peer-to-peer exchange costs $C^{int}_n$ for each member. The peer-to-peer exchange cost for each participant is calculated in a similar way as for the energy exchange between the community and the grid:
\begin{equation} \label{eq:c_int}
    C^{int}_{n}\ =\ \sum_{t\in T}\ (Q^{im}_{t,n}\cdot \lambda_{t,n}^{int,im} - Q^{ex}_{t,n} \cdot \lambda_{t,n}^{int,ex}) \cdot \Delta t
\end{equation}
where internal power flows  ($ Q^{im}_{t,n}$, $ Q^{ex}_{t,n}$) and tariffs ($\lambda_{t,n}^{int,im}$, $\lambda_{t,n}^{int,ex}$) are used. Eq. \ref{eq:int-balance} assures that the power balance inside the community is satisfied, which is based on the community-based market concept presented in \cite{Sousa2019}. Eq. \ref{eq:rec-balance} takes care of the power balance of the internal and external power exchange with the power flows of each member of the REC and Eq. \ref{eq:node-balance} represents the power balance at each users level.
\begin{equation} \label{eq:int-balance}
    \sum_{n \in N}(Q^{im}_{t,n}\ -\ Q^{ex}_{t,n})\ =\ 0
\end{equation}
\begin{equation} \label{eq:rec-balance}
    p_{t,n}\ -\ Q^{im}_{t,n}\ -\ P^{im}_{t,n}\ +\ Q^{ex}_{t,n}\ +\ P^{ex}_{t,n}\ =\ 0
\end{equation}
\begin{equation} \label{eq:node-balance}
    p_{t,n}\ =\ l_{t,n}\ +\ p^{hp}_{t,n}\ +\ p^{cool}_{t,n}\ +\ p^{ev}_{t,n}\ -\ p^{pv}_{t,n}\ +\ p^{ch}_{t,n} \cdot y^{ch}_{t,n}\ -\ p^{disch}_{t,n} \cdot y^{disch}_{t,n}
\end{equation}
With $p_{t,n}$ being the resulting power balance, $l_{t,n}$ the power demand, $ p^{hp}_{t,n} $ the heat demand, $ p^{cool}_{t,n} $ the cooling demand, $ p^{ev}_{t,n} $ the EV charger demand, $ p^{pv}_{t,n} $ the PV power production, $ p^{ch}_{t,n} $ the charging power of the battery and $ p^{disch}_{t,n} $ the discharging power of the battery. 
Heating and cooling loads of the heat pump and electric vehicle all have their own power balance, shown in Eq. \ref{eq:flex-balance}. Their demand $d^{j}_{d,n}$ has to be satisfied on a daily basis by optimizing the hourly usage of the assets taking in consideration the availability $Y^{j}_{i,n}$ of the asset. Eq. \ref{eq:flex-max} ensures that the maximum power of each assets is respected.
\begin{equation} \label{eq:flex-balance}
    \sum_{i \in I^d}\ (p^{j}_{i,n} \cdot Y^{j}_{i,n})\ =\ d^{j}_{d,n}
\end{equation}
\begin{equation} \label{eq:flex-max}
    p^{j}_{t,n}\ \leq\ m^{j}_{n} 
\end{equation}
Eq. \ref{eq:pv} sets the power output of the PV installation based on the normalized electricity production from PV $G_t$ and its capacity $cap_{pv,n}$.
\begin{equation} \label{eq:pv}
    p_{t,n}^{pv}\ =\ cap_{pv,n}\cdot G_{t}
\end{equation}
Eq. \ref{eq:ess} represents the energy balance of the battery: $ e_{t,n} $ is the energy content of the BESS, $\eta^{ch}$ its charging efficiency and $\eta^{disch}$ its discharging efficiency. The inclusion of binary variables $y^{ch}_{t,n}$ and $y^{disch}_{t,n}$ ensures that the battery is not charging and discharging at the same time (Eq. \ref{eq:ess-binary}).
\begin{equation} \label{eq:ess}
    e_{t+1, n}\ =\ e_{t,n}\ +\ \Delta t \cdot (y^{ch}_{t,n} \cdot p^{ch}_{t,n} \cdot \eta^{ch}\ -\ y^{disch}_{t,n} \cdot p^{disch}_{t,n}/\eta^{disch})
\end{equation}
\begin{equation} \label{eq:ess-binary}
    y^{ch}_{t,n}\ +\ y^{disch}_{t,n}\ \leq\ 1
\end{equation}
In addition boundaries for the charging and discharging powers of the battery based on maximum C-rates are introduced:
\begin{equation} \label{eq:charge}
   p^{ch}_{t,n}\ \leq\ \overline{p^{ch}} \cdot cap_{batt,n}
\end{equation}
\begin{equation} \label{eq:discharge}
   p^{disch}_{t,n}\ \leq\ \overline{p^{disch}} \cdot cap_{batt,n}
\end{equation}
as well as a minimum and maximum state-of-charge:
\begin{equation} \label{eq:socminmax}
    SOC_{min} \cdot cap_{batt,n}\ \leq\ e_{t,n}\ \leq\ cap_{batt,n}
\end{equation}
\section{Results}
\label{sec:results}
Results are presented in boxplots in order to make the comparison between the 156 scenarios more readable. To interpret the results one needs to compare the same points between a group of boxplots (minimum with minimum, median with median, etc.), as such points represent the same set of scenarios. 
Three of the scenarios' categories results - capex, tariff and investment - are  presented in separate groups of boxplots in order to assess their individual impact on the KPIs. Additionally, the variation between different scenarios in the technology scenario set has a large impact on the KPIs. This is the reason why when presenting the results, the technology scenario set of 4 scenarios are presented alongside each of the three previously mentioned scenario sets. Finally, internal price scenarios apply only within the third investment scenario, the REC.
\subsection{CAPEX scenarios results}
\begin{figure}[h!]
    \centering
    \includegraphics[width=.95\linewidth]{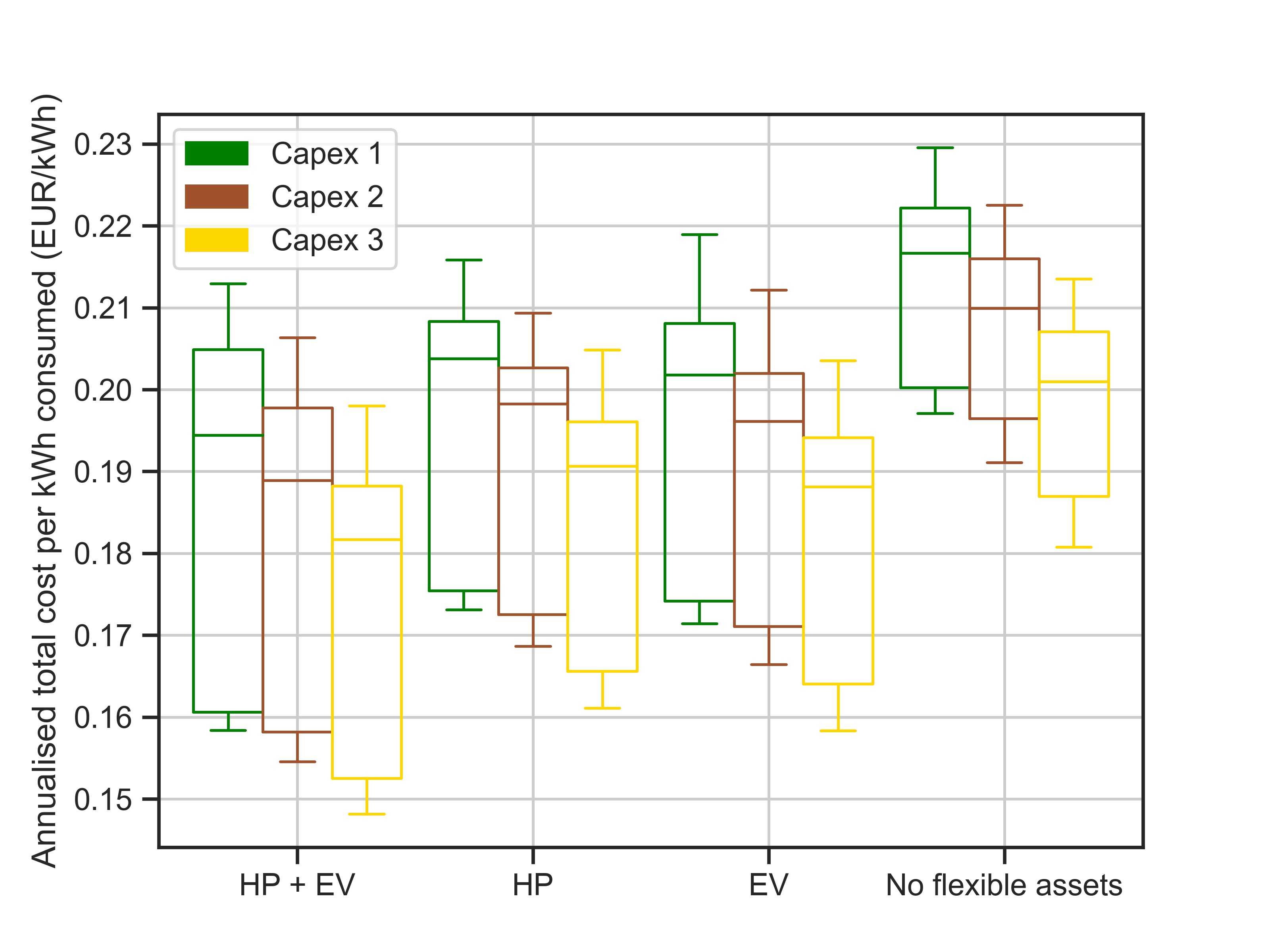}
    \caption{Total annualized cost per kWh - Capex scenarios comparison}
    \label{fig:tac-capex}
\end{figure}

The comparison between different CAPEX for PV and BESS on the total annualized cost is shown in Figure \ref{fig:tac-capex}. Predictably, lower investment cost results in lower total cost of electricity, but their impact is lower compared to the one of the different electricity tariffs. 
Moreover, lower prices for assets lead to higher PV capacities installed (see Figure \ref{fig:pv-per-MWh-capex}), and also in this case we can see that the optimal size of PV increase with the electrification of heating and transportation. BESS will benefit even more than PV of a reduction of investment cost because current prices (CAPEX scenario 1) will result almost every time in non installing any storage system as optimal choice (see Figure \ref{fig:bess-per-MWh-capex}) and BESS become more cost-efficient in absence of flexibility in the system. As seen in previous scenarios, higher installation of PV (Capex scenario 3) will decrease self-consumption ratio (Figure \ref{fig:scr-capex}) and increase self-sufficiency ratio (Figure \ref{fig:ssr-capex}). Increase of BESS capacities, which also happen mostly for the third Capex scenario, should theoretically increase both KPIs. However, their capacities are relatively smaller than PV ones and therefore they do not have a large impact on the results.
\subsection{Tariff scenarios results}
\begin{figure}[h!]
    \centering
    \includegraphics[width=.95\linewidth]{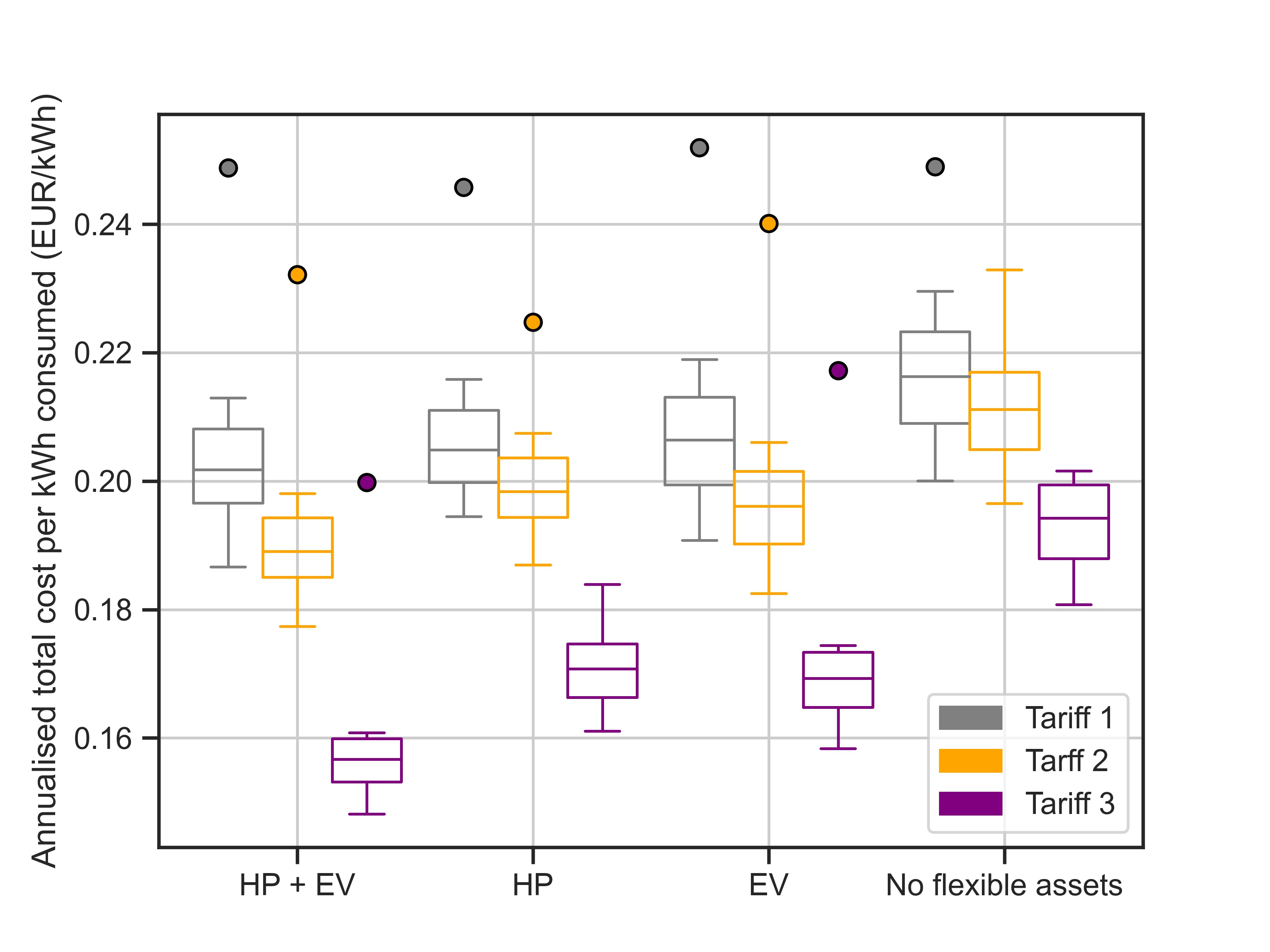}
    \caption{Total annualized cost per kWh - Tariff scenarios comparison}
    \label{fig:tac-tariff}
\end{figure}

The total annualized cost per kWh consumed compared by tariff is shown in Figure \ref{fig:tac-tariff}, where we can see that increasing the amount of capacity-dependent component in the electricity tariff will reduce electricity cost. The smaller variability between results in each tariff scenario compared to other scenario comparison (investment and technology scenarios), indicates that tariffs are one of the main parameter affecting the results.
Regarding PV installation with different tariffs, we observe that capacity tariffs will decrease the optimal capacity installed (see Figure. \ref{fig:PV-per-MWh-tariff}). This confirms that the size of the PV installation is proportional to the total electricity demand as the capacity installed per consumption unit is very similar for each technology scenario. For what it concerns batteries installation, we can see that the optimal installed capacities follow a different tendency than what was seen for PV: BESS are chosen almost only when a capacity tariff is implemented (see Figure \ref{fig:BESS-per-MWh-tariff}). For the investment scenarios, a lack of flexibility in the system will also increase the amount of BESS installed.
Capacity tariffs also increase self-consumption compared to more volumetric tariffs (Figure \ref{fig:SCR-tariff}) due to less PV and more BESS present in the system. On the other hand, for the same reasons, self-sufficiency ratio decreases when the tariff is moving towards a capacity-based one (Figure \ref{fig:SSR-tariff})
\subsection{Investment scenarios results}
\begin{figure}[ht!]
    \centering
    \includegraphics[width=.95\linewidth]{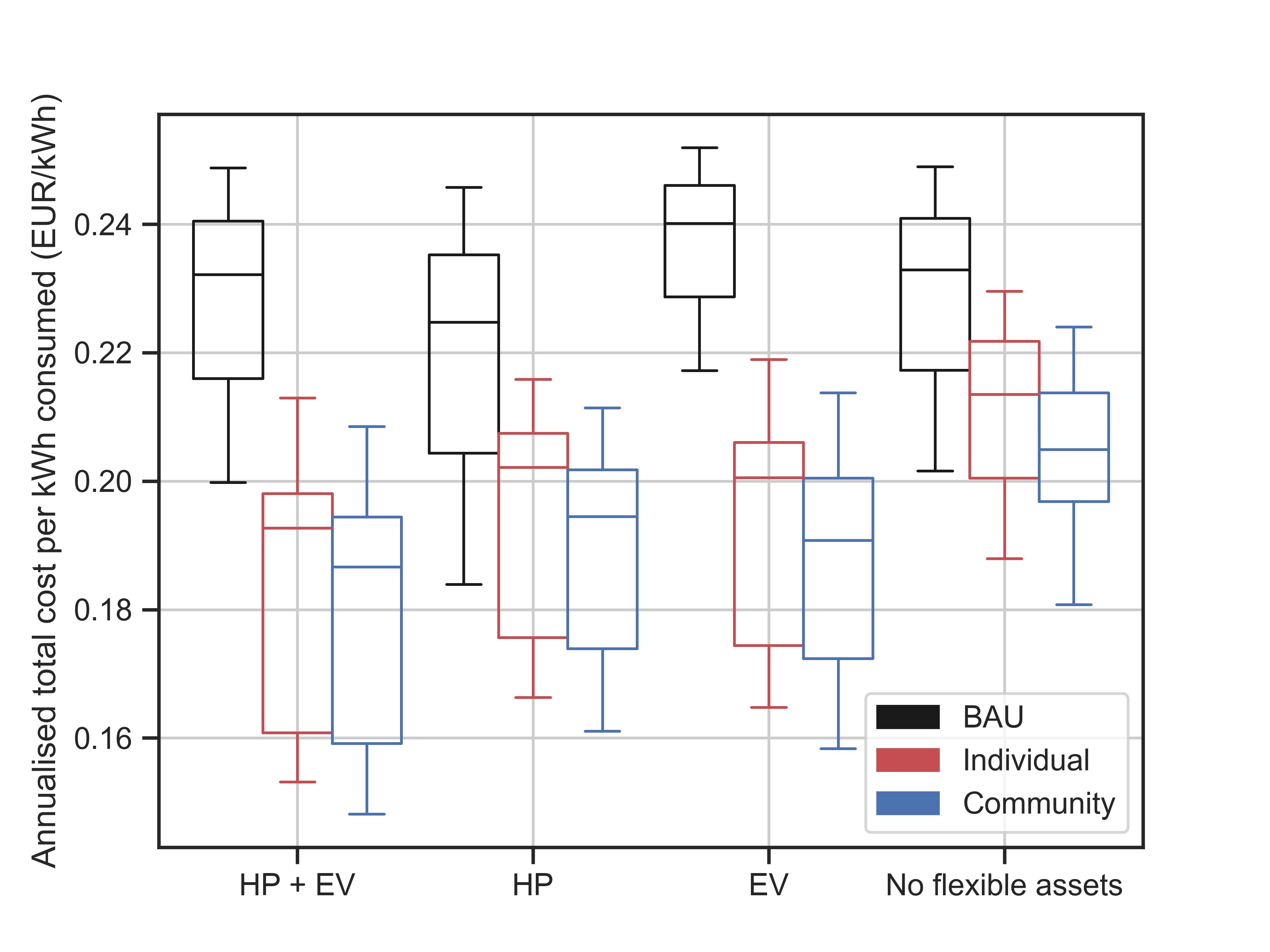}
    \caption{Total annualized cost per kWh - Investment scenarios comparison}
    \label{fig:TAC-per-kWh}
\end{figure}

In general, independently of the technology scenarios the community investment option always results in a better economic result. As presented in Figure \ref{fig:TAC-per-kWh} in the case where investments are made, having more flexible assets (hence, higher electricity consumption) leads to lower electricity cost per kWh consumed. It also can be seen, that i) the variability of results (height of the boxes) increase with higher penetration of flexible assets and ii) investing in PV and BESS leads to a lower annualized total cost per kWh compared to the BAU scenarios.
PV capacity installed per consumption unit stays very similar for each technology scenario, i.e. a higher electricity consumption (HP + EV scenario) will increase the PV capacity installed for optimal solutions (see Figure \ref{fig:PV-inv-per-MWh}). It should be also noted that community investment scenarios allow the highest PV and BESS integration. However, the cost-efficiency of PV and BESS follow different trends in relation with the amount of existent flexibility in the system. Higher capacity of BESS is more cost-efficient when less flexibility is already available in the system (see Figure \ref{fig:ESS-inv-per-MWh}) while the opposite applies for PV. Concerning the last two KPIs, the self-consumption ratio balances out between technology scenarios (see Figure \ref{fig:SCR-inv}) while the self-sufficiency ratio increases due to higher flexibility assets and PV penetration (see Figure \ref{fig:SSR-inv}). For both KPIs, the community scenarios give the best results.
\subsection{Internal price scenarios results}
\begin{figure}[ht!]
    \centering
    \includegraphics[width=.95\linewidth]{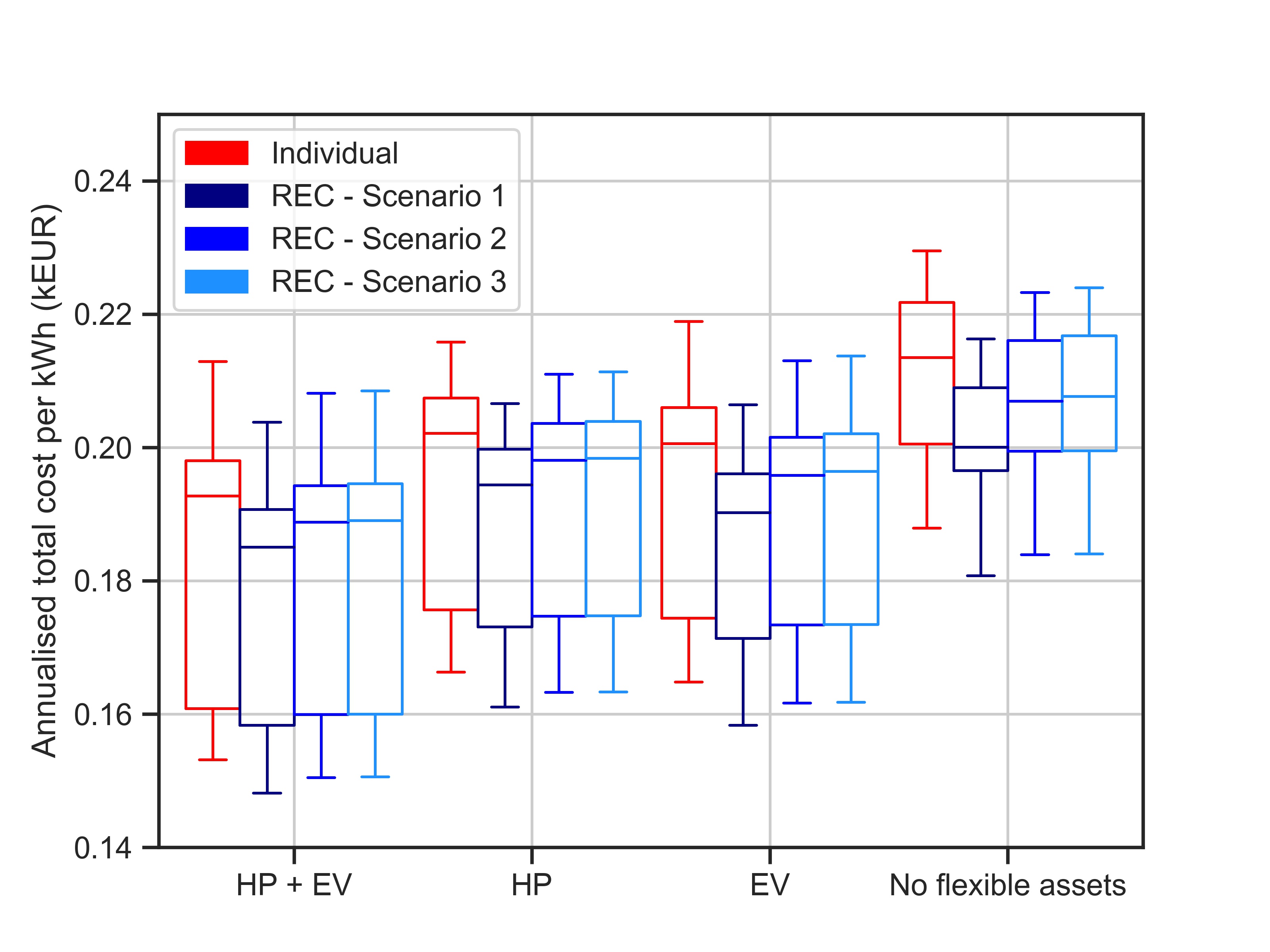}
    \caption{Internal cost impact}
    \label{fig:internal-cost}
\end{figure}

The impact on varying the internal energy exchange  price is shown in Figure \ref{fig:internal-cost}. We can see how a higher cost for internal exchange will reduce the already small gain, maximum 6 \% at best on the annualized cost per kWh consumed, of the REC over the individually optimized buildings.
\section{Discussion}
\label{sec:discussion}
Investments in DER such as PV and storage systems such as batteries generate a positive business case for the users. The creation of a REC always results in a cost reduction compared to the reference case, with a reduction varying from 10\% to 26 \%. Similar numbers were found in an equivalent study in the Austrian context \cite{Cosic2021}, where the creation of a REC becomes eventually the economical best solution for all the scenarios included. However, the economic advantage only slightly increase with the creation of a REC compared to individual investment and operation. RECs are able to decrease the annualised total cost per energy consumed of at maximum 6 \% compared to single user of an optimized building. This is mainly achieved by introducing the possibility to exchange energy between prosumers. On top of this one also needs to considers the IT and IoT costs which can completely erase this small gain. Subsidies might help but can create unjust societal situations that have to be studied from an entire system perspective. \par
On site DER production for countries with such a high population density and low levels of direct solar radiation, like Belgium, can expect limitations for RECs related to be a community of place (where there is a proximity constraint) and not of interest (where the proximity to the energy generation assets does not necessarily play a role). This will limit technological options for renewable self production of energy to PV. Moreover, reaching high levels of self-sufficiency and self-consumption requires high levels of electrification of multiple energy vectors, like heat and transport.\par
With the electricity prices for Flanders, the switch towards a larger power component in the electricity bill results in a lower final cost for the users, especially in the presence of optimal control of BESS or a HP and EV charger. This is achieved by the ability to reduce the peak consumption, which is also a positive outcome for the network operator. However, our findings show that in this situation it is more convenient to install smaller capacities of PV compared to traditional volumetric tariffs. This reduces the potential of RECs to add RES generation capacity and limits their impact in the decarbonsation of the energy system. On the other hand, a move towards a more capacity-based tariff will trigger more investment into BESS, as found by Radl et al. \cite{Radl2020} for other European countries. In general, a more dynamic pricing mechanism would be beneficial for RECs, as highlighted by \cite{Fernandez2021}, where the higher profit is reached when real-time pricing is used.\par
One of the main limitation of the optimization model developed in this study is that it works with a perfect foresight of the input profiles - consumption and solar radiation - which take out all the real-life uncertainties. Possible future work could be the introduction of stochastic input parameters to further handle these uncertainties. Another difficulty encountered in this work is to create a general model for REC because it could have all sorts of prosumers/consumers and regulatory conditions. The consultation process whit stakeholders allowed us to include typical diversity that can be found in Flanders but efforts have to be invested in the creation of typologies of RECs, which would allow the diversity necessary to conduct system wide analysis.
\section{Conclusions}
\label{sec:conclusions}
The creation of a REC is always outperforming the other scenarios in both economical and renewable energy penetration KPIs. However, there is never a substantial economical advantage for REC over individual smart-houses with own electricity generation assets. Moreover, there are still a lot of uncertainties on the regulation of REC in Flanders and how peer-to-peer exchange will work, which could potentially shift the benefits towards one scenario or the other. In addition to that, the cost for setting up a REC, both administrative and technical costs, are not included in this or any other similar analysis. Hence, the small gain of constituting a REC could potentially be erased by these additional costs. The shift towards a capacity tariff would help the uptake of BESS as they will become more cost-efficient due to their ability to shift peak demand. This will correspond to less PV installed to reach the optimal situation compared to volumetric tariff. Regarding the total cost comparison between volumetric and capacity-based tariff, the latter gives the best results and in general are one of the two most influential factors on the final cost of electricity. The other very impactful factor on the annualised cost of energy is the amount of flexibility present in the system. In fact, electrification of heat and transport will add more possibilities to reduce costs in a smart energy system. The consequent increase of electrical demand will also make PV installations more cost-efficient with respect to situation where heat and transport are not powered by electricity.

\section*{Acknowledgements}
\label{sec:acknowledgement}
This project has received funding from the European Union's Horizon 2020 research and innovation program under grant agreement No 824342 as well as from VLAIO in the ICON project ROLECS (reference HBC.2018.0527) and the project MAMUET (grant number HBC.2018.0529).

\section{CRediT authorship contribution statement}
\textbf{Alex Felice:} Conceptualization, Methodology,  Validation, Formal analysis, Investigation, Data Curation, Software, Writing - Original Draft, Writing - Review \& Editing, Visualization \textbf{Lucija Rakocevic:} Conceptualization, Methodology, Writing - Review \& Editing. \textbf{Leen Peeters:} Resources, Project administration, Funding acquisition. \textbf{Maarten Messagie:} Resources, Project administration, Funding acquisition. \textbf{Thierry Coosemans:} Resources, Project administration, Funding acquisition.
\textbf{Luis Ramirez Camargo:} Conceptualization, Methodology, Writing - Review \& Editing.
\bibliographystyle{elsarticle-num}
\typeout{}
\bibliography{alex_energy_communities_paper}





\appendix

\section{Supplementary material}
\label{sec:sample:appendix}
\begin{figure}[p]
    \centering
    \includegraphics[width=.9\linewidth]{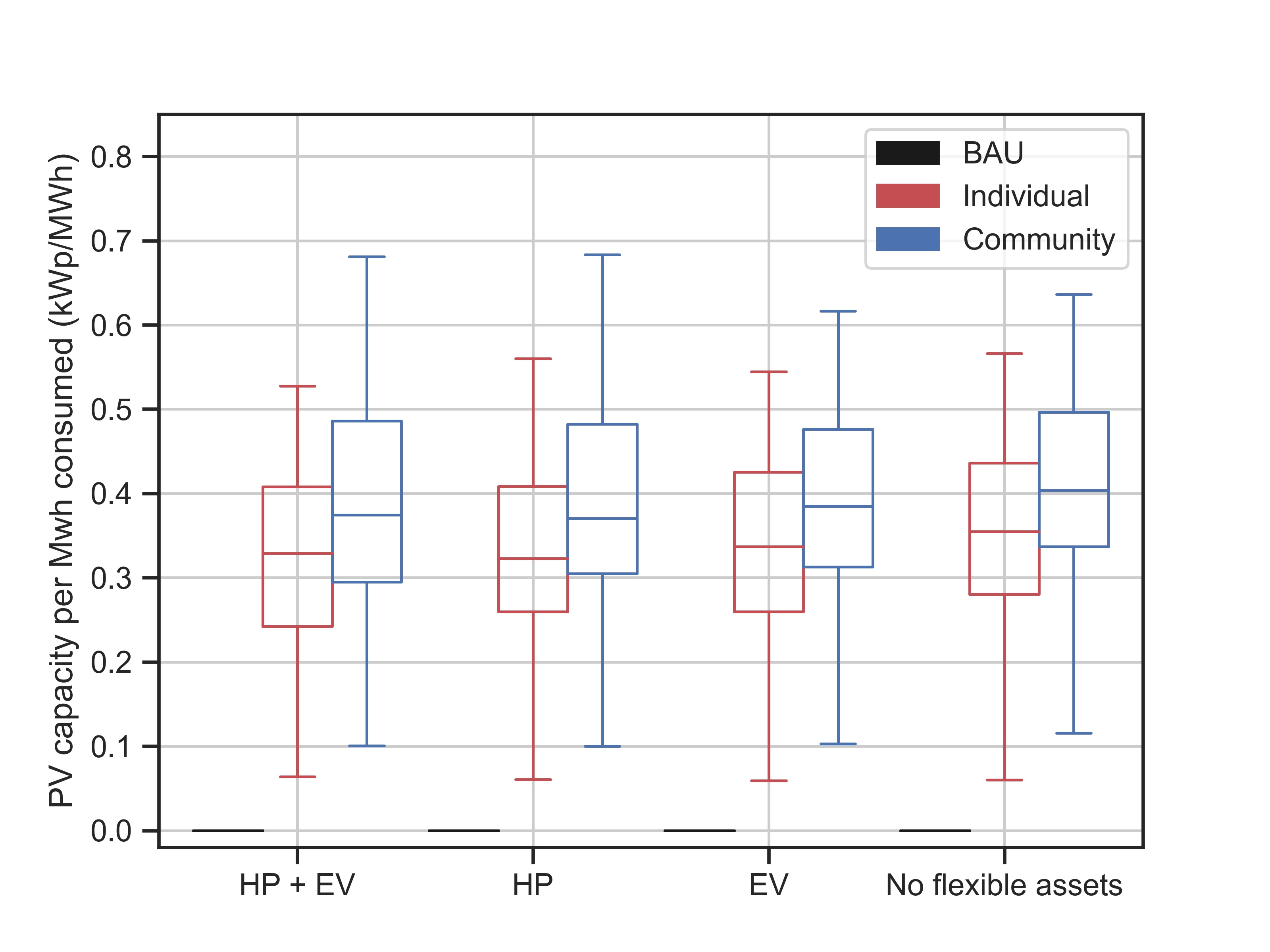}
    \caption{PV capacity installed per MWh consumed - Investment scenarios comparison}
    \label{fig:PV-inv-per-MWh}
    \includegraphics[width=.9\linewidth]{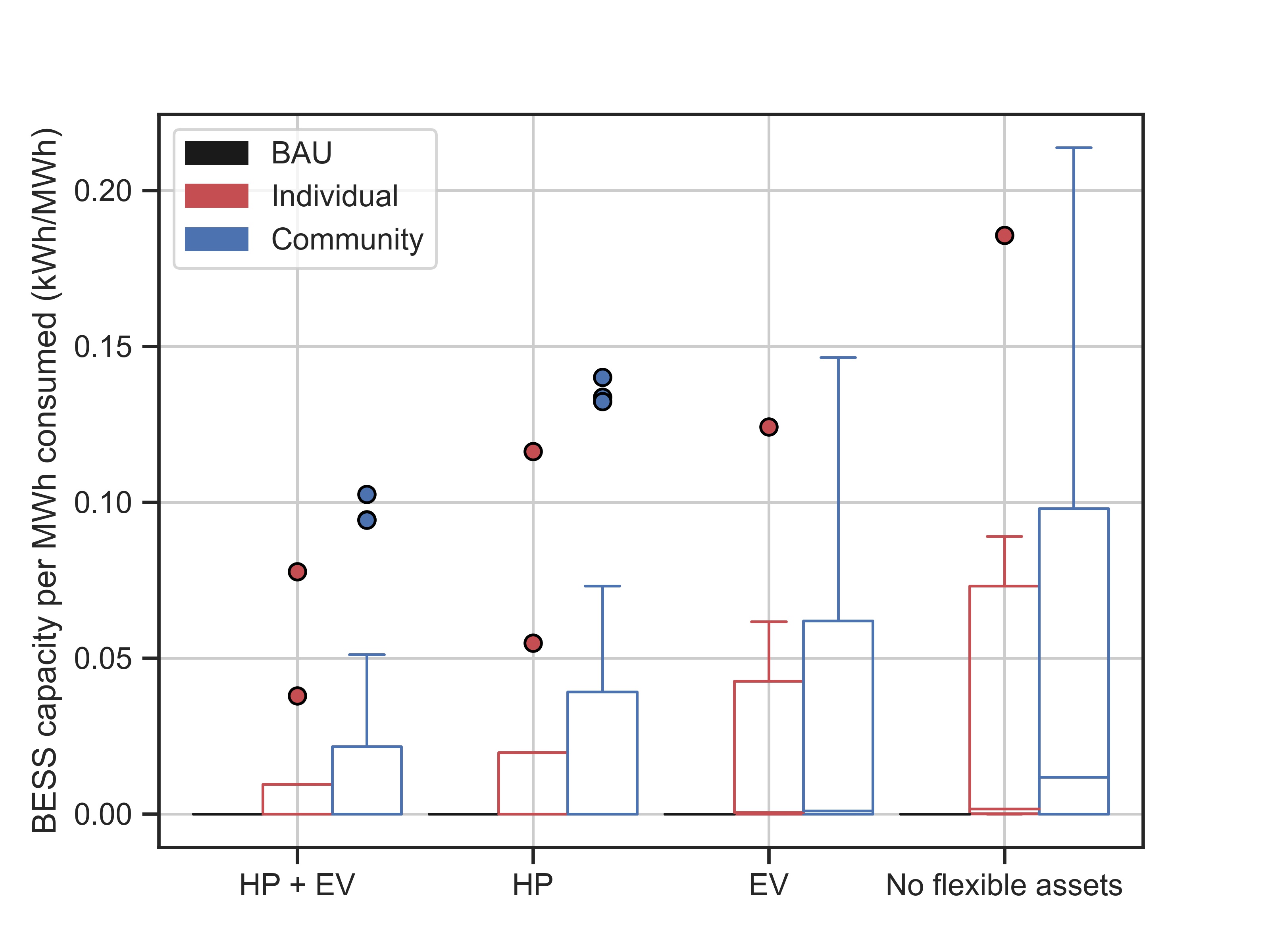}
    \caption{BESS capacity installed per MWh consumed - Investment scenarios comparison}
    \label{fig:ESS-inv-per-MWh}
\end{figure}
\begin{figure}[p]
    \centering
    \includegraphics[width=.9\linewidth]{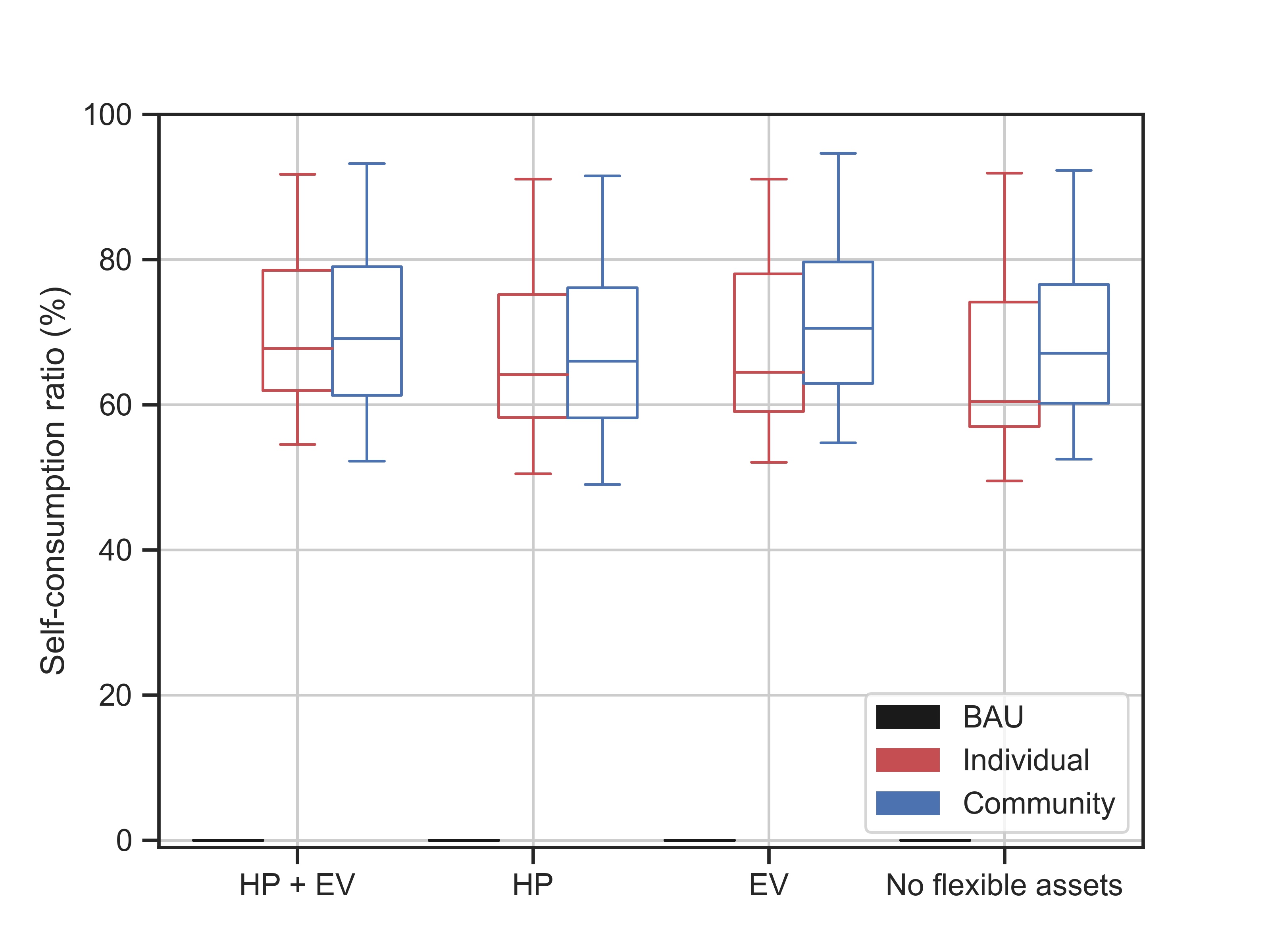}
    \caption{SCR - Investment scenarios comparison}
    \label{fig:SCR-inv}
    \includegraphics[width=.9\linewidth]{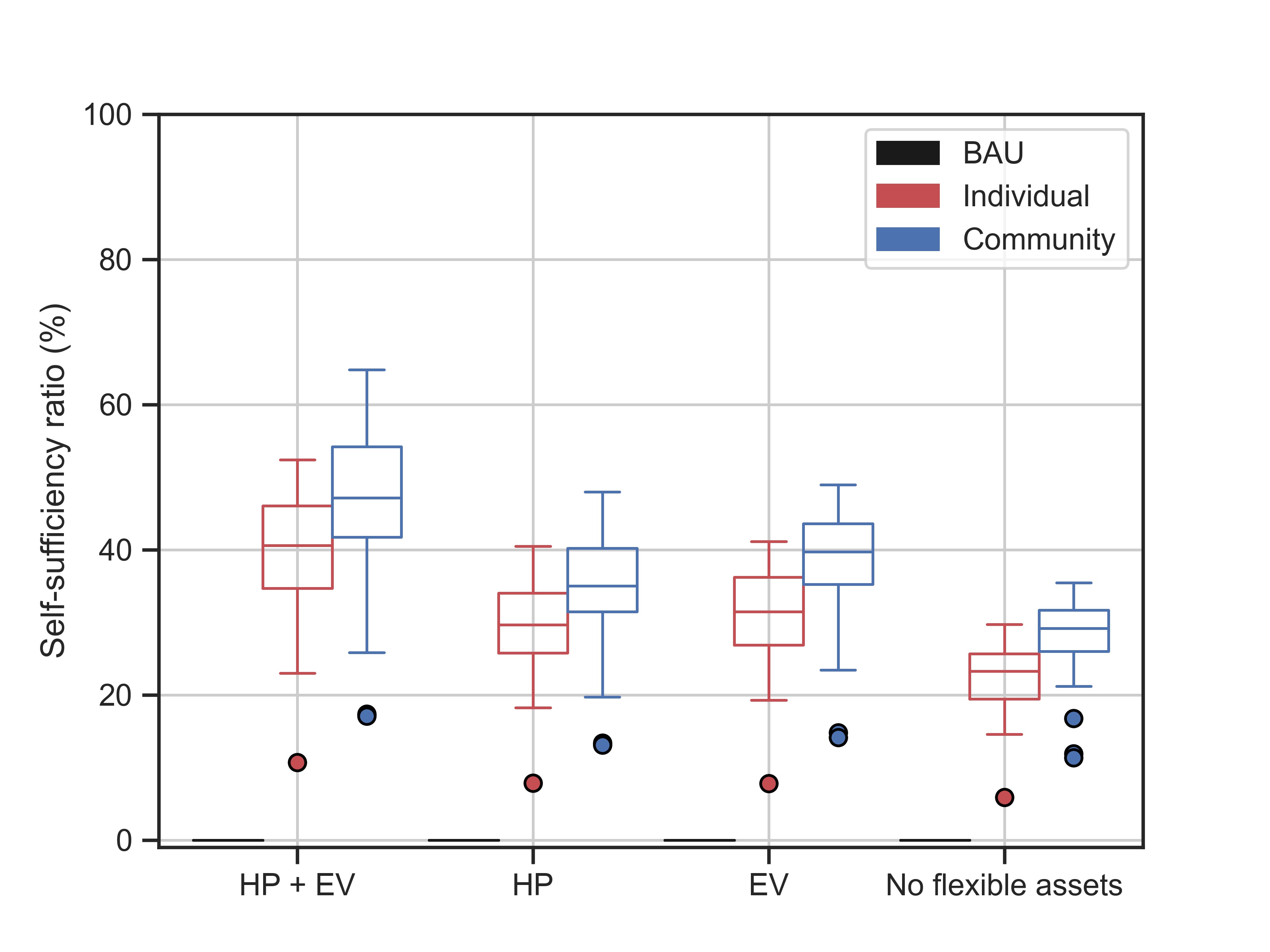}
    \caption{SSR - Investment scenarios comparison}
    \label{fig:SSR-inv}
\end{figure}
\begin{figure}[p]
    \centering
    \includegraphics[width=.9\linewidth]{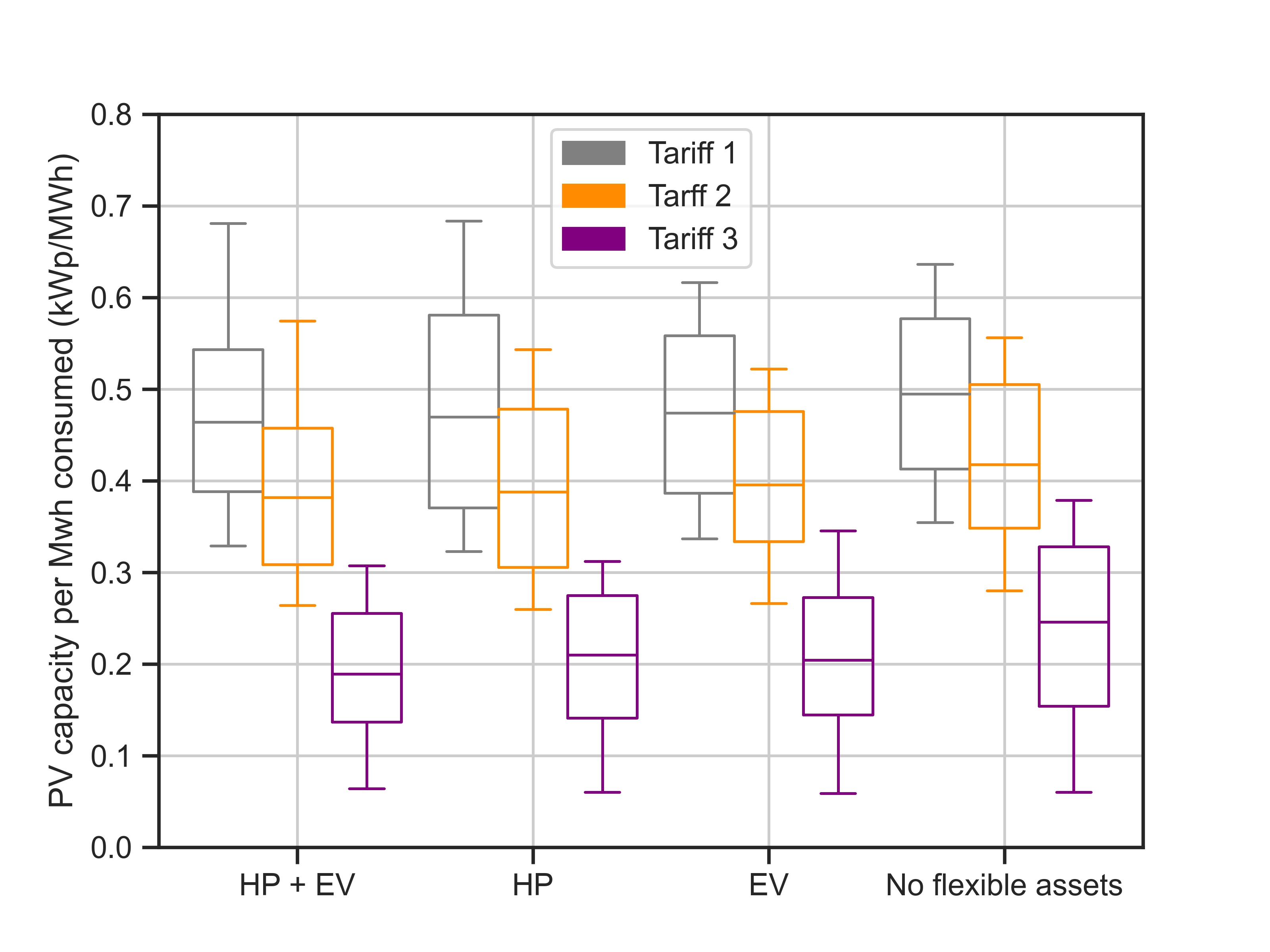}
    \caption{PV capacity installed per MWh consumed - Tariff scenarios comparison}
    \label{fig:PV-per-MWh-tariff}
    \includegraphics[width=.9\linewidth]{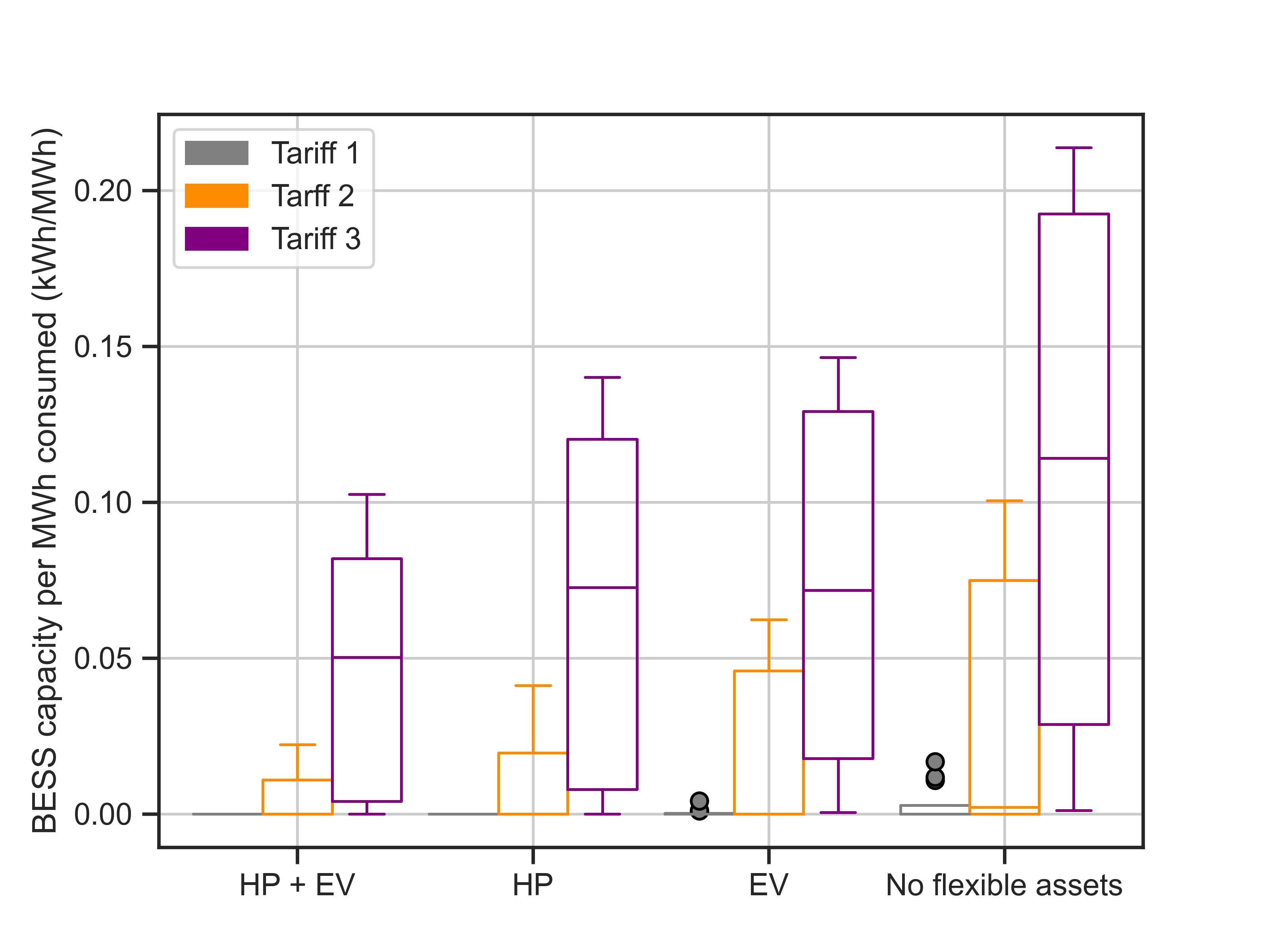}
    \caption{BESS capacity installed per MWh consumed - Tariff scenarios comparison}
    \label{fig:BESS-per-MWh-tariff}
\end{figure}
\begin{figure}[p]
    \centering
    \includegraphics[width=.9\linewidth]{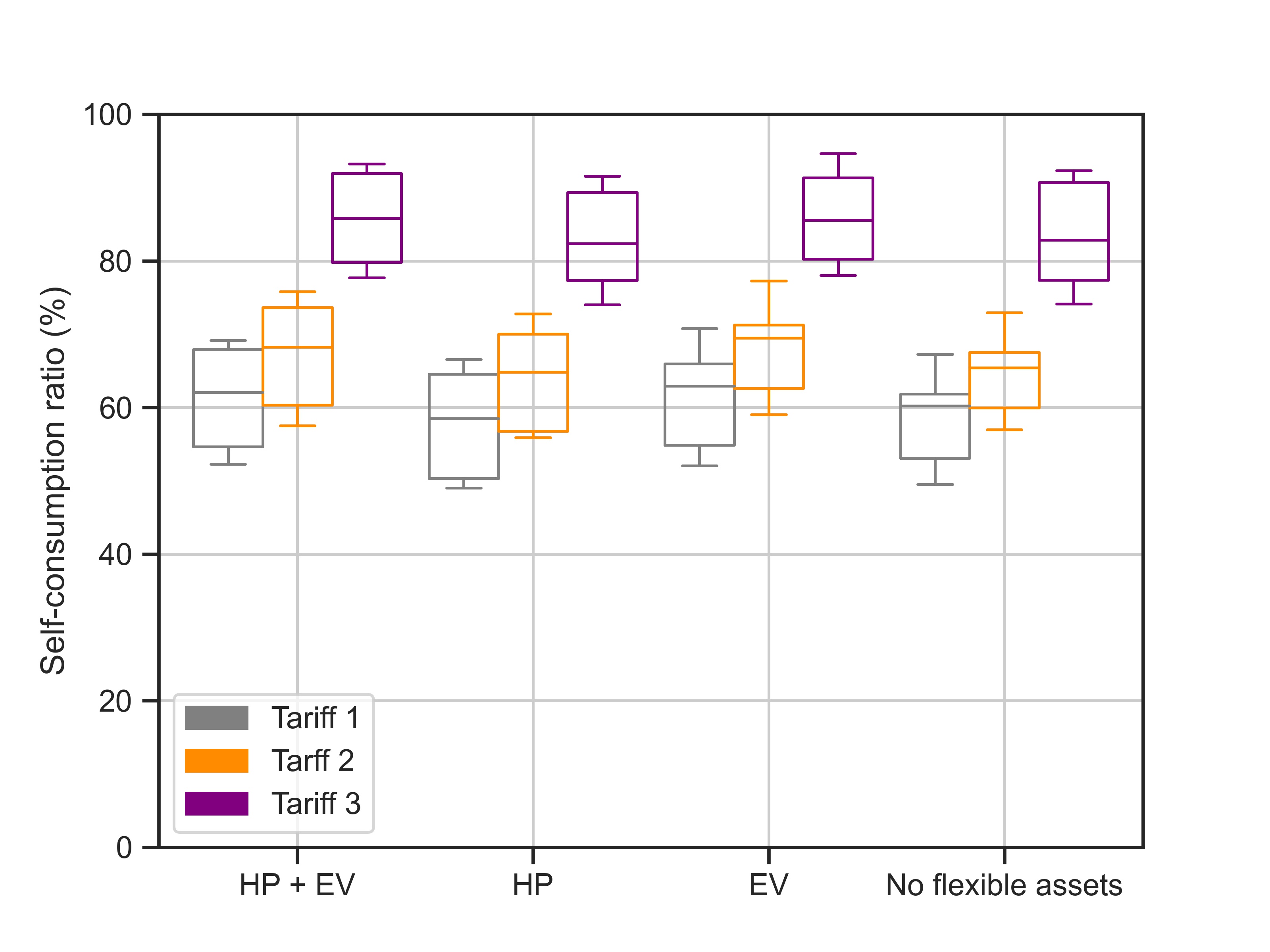}
    \caption{SCR - Tariff scenarios comparison}
    \label{fig:SCR-tariff}
\end{figure}
\begin{figure}[ht!]
    \centering
    \includegraphics[width=.9\linewidth]{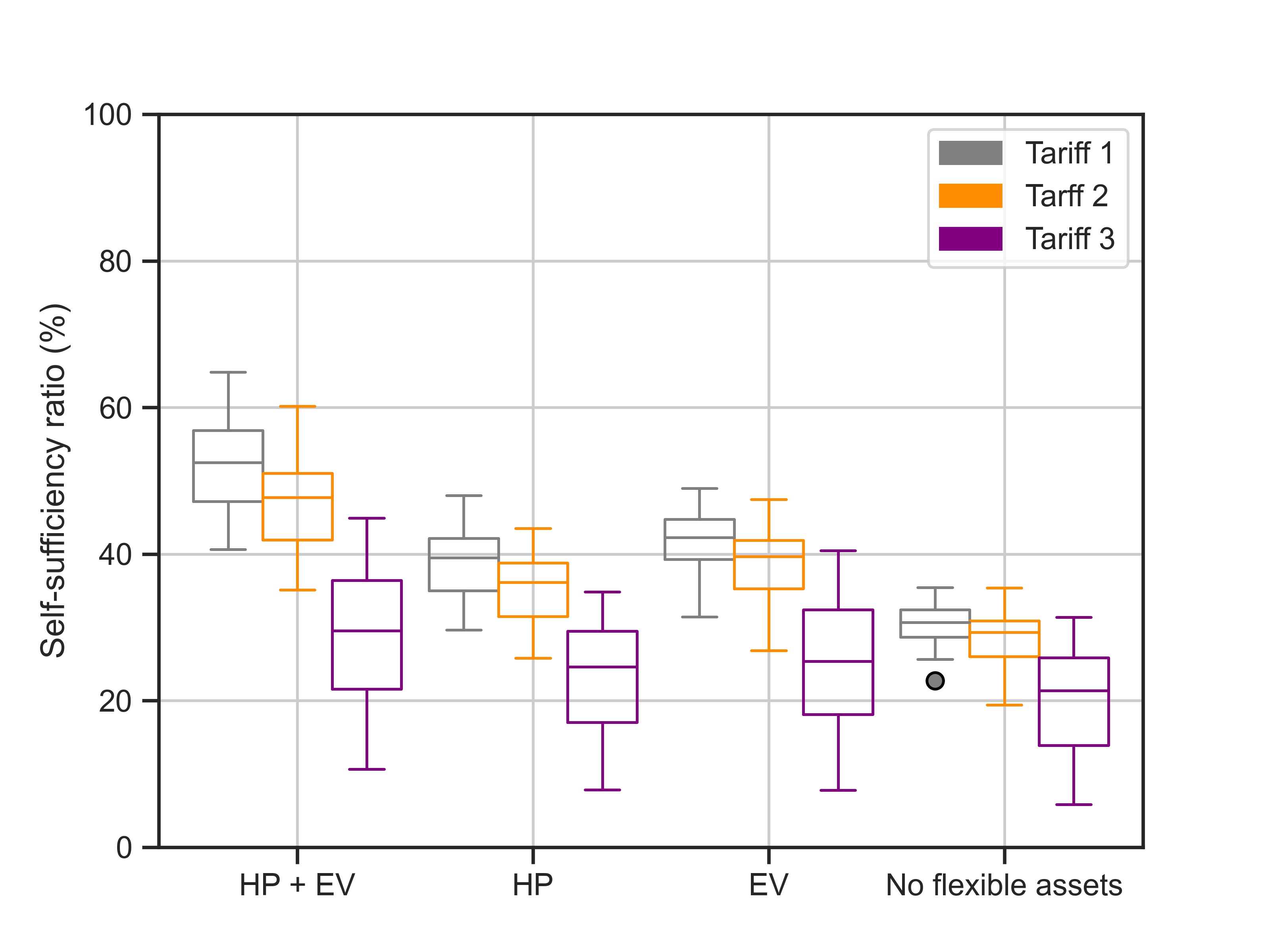}
    \caption{SSR - Tariff scenarios comparison}
    \label{fig:SSR-tariff}
\end{figure}
\begin{figure}[p]
    \centering
    \includegraphics[width=.9\linewidth]{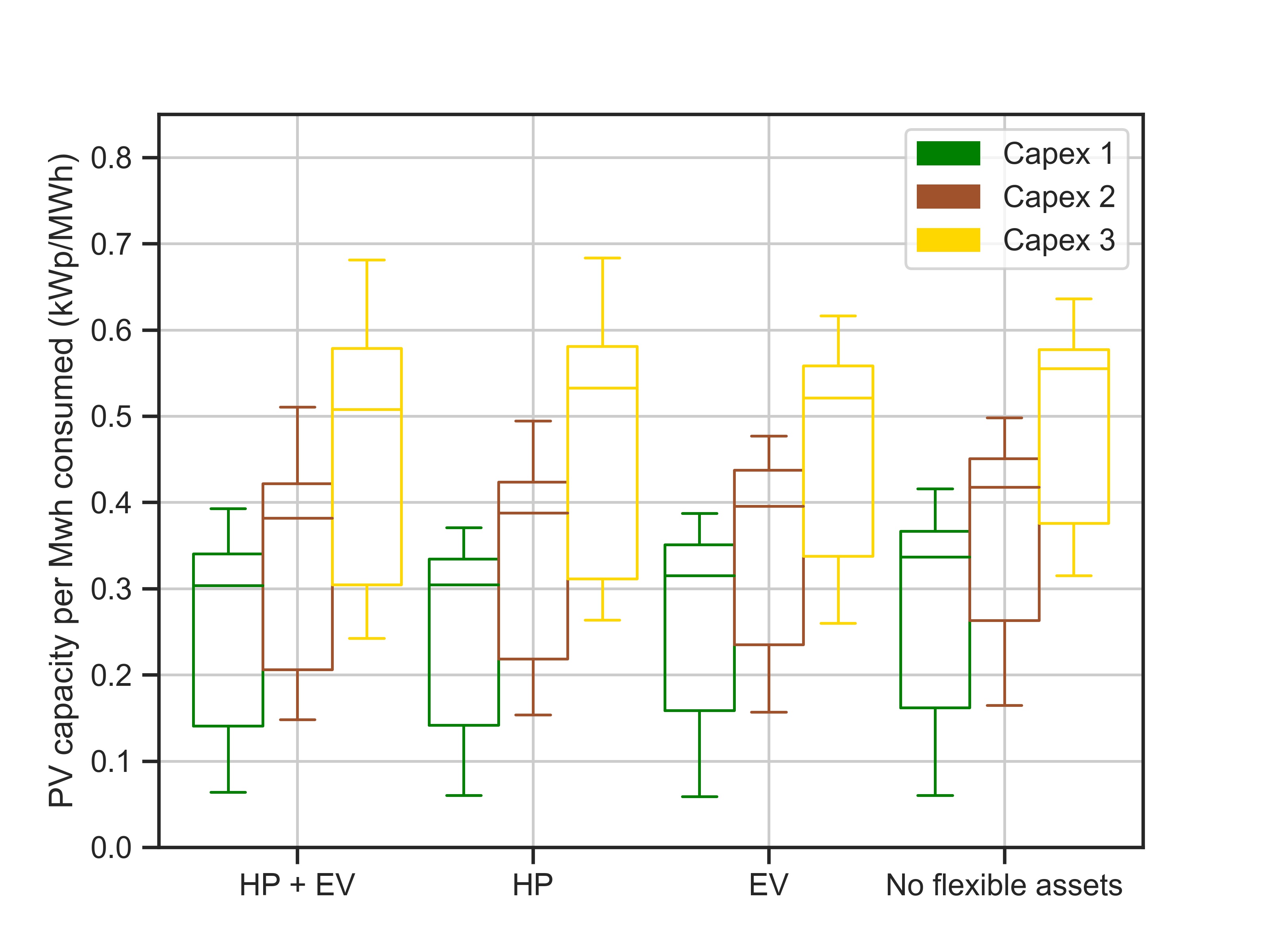}
    \caption{PV capacity installed per MWh consumed- Capex scenarios comparison}
    \label{fig:pv-per-MWh-capex}
    \includegraphics[width=.9\linewidth]{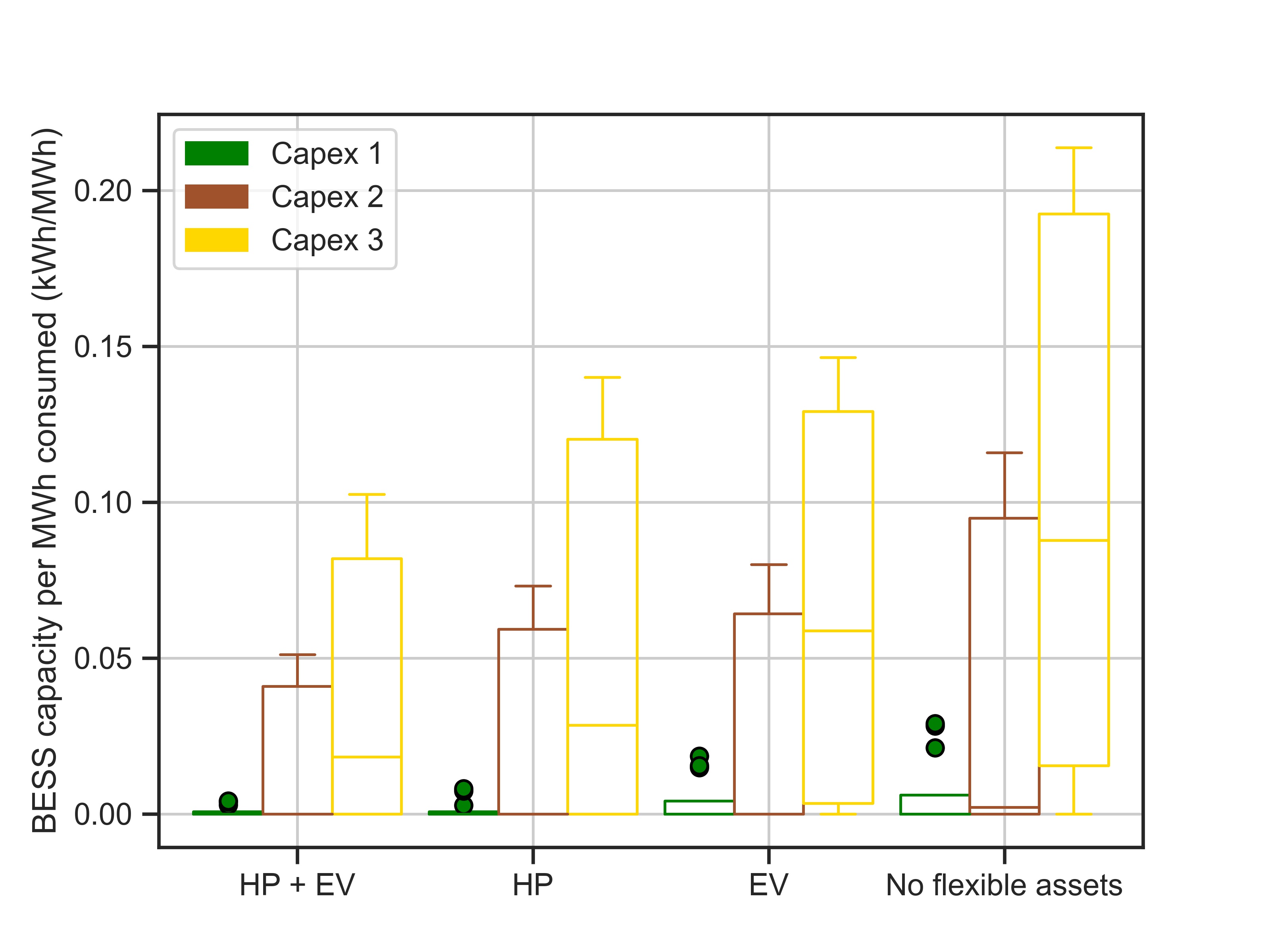}
    \caption{BESS capacity installed per MWh consumed - Capex scenarios comparison}
    \label{fig:bess-per-MWh-capex}
\end{figure}
\begin{figure}[p]
    \centering
    \includegraphics[width=.9\linewidth]{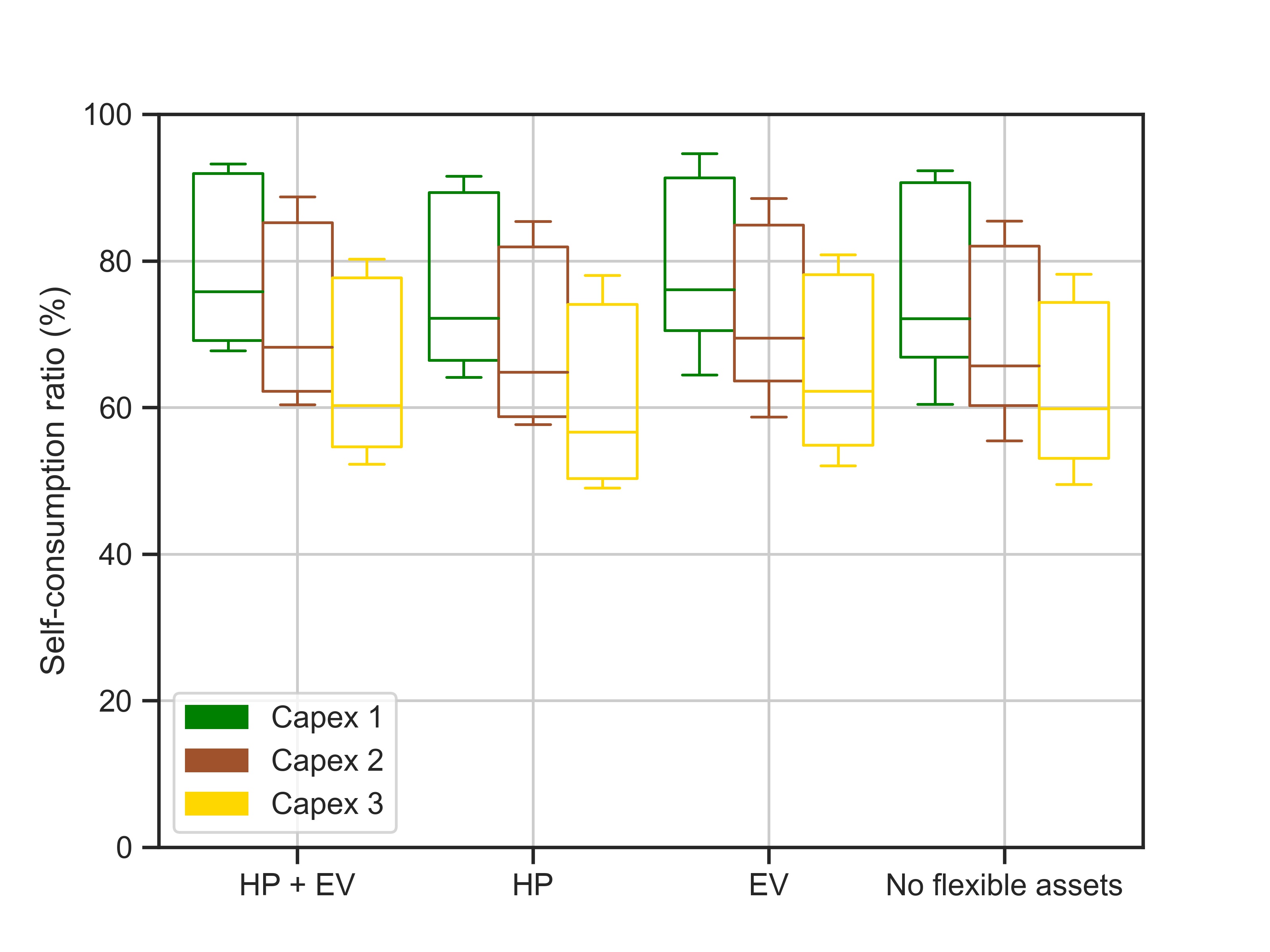}
    \caption{SCR - Capex scenarios comparison}
    \label{fig:scr-capex}
    \includegraphics[width=.9\linewidth]{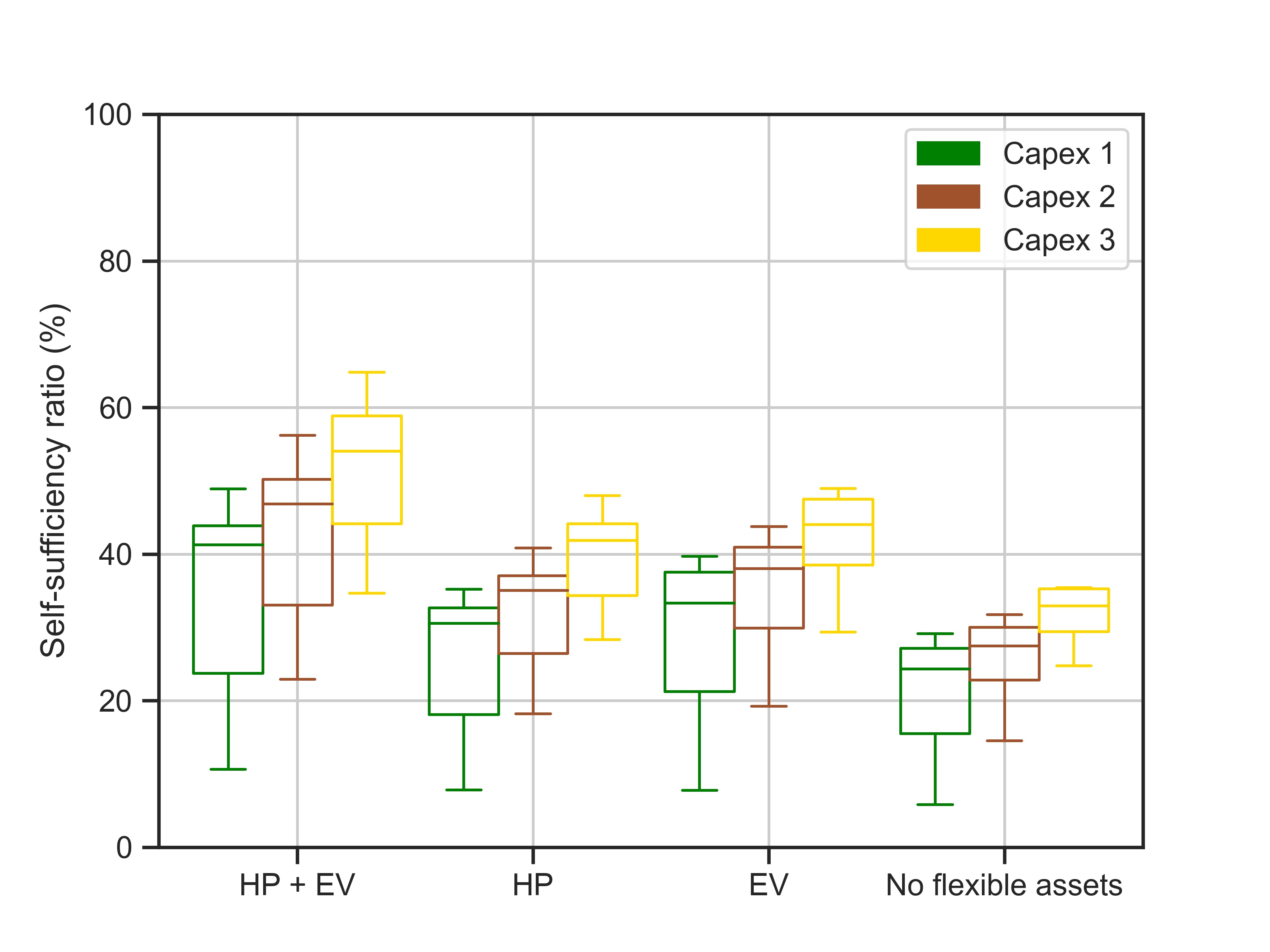}
    \caption{SSR - Capex scenarios comparison}
    \label{fig:ssr-capex}
\end{figure}

\end{document}